\documentclass[journal]{IEEEtran}

\usepackage{amsmath,epsfig,amssymb,verbatim,amsopn,citesort,subfigure,color,multirow}

\newcommand{\Pt}{\mathcal{P}}

\newcommand{\Ptx}{{p_{\text{tx}}}}
\newcommand{\Pins}{{p_{\text{so}}}}

\newcommand{\se}{\sigma^2_{e}}
\newcommand{\sn}{\sigma^2_{b}}

\newcommand{\Gb}{\bar{\gamma}_b}
\newcommand{\Ge}{\bar{\gamma}_e}
\newcommand{\Gt}{\mu}
\newcommand{\Gi}{\bar{\gamma}_i}

\def\onedot{.,\,}

\def\ie{i.e\onedot}

\newcommand{\AuthorOne}{Xiangyun Zhou}
\newcommand{\AuthorThree}{Behrouz Maham}
\newcommand{\AuthorTwo}{Matthew R. McKay}
\newcommand{\AuthorFour}{Are Hj{\o}rungnes}

\newcommand{\ThankOne}{X. Zhou, B. Maham, and A. Hj{\o}rungnes are with UNIK - University Graduate Center, University of Oslo, Kjeller, Norway. (Email: \{xiangyun,behrouz,arehj\}@unik.no). M. McKay is with the Department of Electronic and Computer Engineering, Hong Kong University of Science and Technology, Hong Kong (Email: eemckay@ust.hk). This work was supported by the Research Council of Norway through the projects 197565/V30 and
176773/S10.}

\title{\LARGE{Rethinking the Secrecy Outage Formulation: A Secure Transmission Design Perspective}}

\author{
\authorblockN{\AuthorOne, \textit{Member, IEEE}, \AuthorTwo, \textit{Member, IEEE}, \AuthorThree, \textit{Member, IEEE}, \\
and \AuthorFour, \textit{Senior Member, IEEE}
\thanks{\ThankOne}
\vspace{-3mm}
}}

\begin{document}

\maketitle

\begin{abstract}

This letter studies information-theoretic security without knowing the eavesdropper's channel fading state. We present an alternative secrecy outage formulation to measure the probability that message transmissions fail to achieve perfect secrecy. Using this formulation, we design two transmission schemes that satisfy the given security requirement while achieving good throughput performance.

\end{abstract}


\vspace{-1mm}

\section{Introduction}

The problem of achieving information-theoretic security over wireless channels without knowing the eavesdropper's instantaneous channel state information (CSI) has attracted considerable attention recently. The performance limits of such systems have been characterized in terms of the ergodic secrecy capacity (see, e.g.,~\cite{gopala_08,khisti_09a}), which captures the capacity limits under the constraint of perfect secrecy. This measure applies for delay-tolerant systems, assuming that the encoded messages span sufficient channel realizations to capture the ergodic features of the channel. For systems with stringent delay constraints, however, perfect secrecy cannot always be achieved, and outage-based characterizations become more appropriate by providing a probabilistic performance measure of secure communication.

The idea of secrecy outages has been adopted in~\cite{parada_05,bloch_08}, where an outage formulation was presented to capture the probability of having a transmission that is both reliable and secure~\cite{parada_05,bloch_08,goel_08,vilela_10}. As we will discuss, however, that formulation does not give a direct indication of the security level in the system, since it accounts for outage events which do not necessarily reflect failures in achieving perfect secrecy. In this letter, we put forth an alternative formulation which gives a more direct measure of system security, accounting for design parameters such as the rate of the transmitted codewords as well as the condition under which transmission is suspended. This new formulation provides a useful framework for computing the probability that transmitted messages are perfectly secure against eavesdropping. Based on our formulation, we also design two transmission schemes, each of which guarantees a certain level of security whilst maximizing the throughput. The first scheme requires CSI feedback from the legitimate receiver to the transmitter, whereas the second scheme requires only $1$ bit of feedback.

\vspace{-1mm}

\section{System Model} \label{sec:}

We consider the transmission of confidential messages from Alice to Bob over a Rayleigh fading channel in the presence of an eavesdropper, Eve. The transmit power, $\Pt$, is assumed fixed to the maximum level. The channel gains from Alice to Bob and Eve are denoted as $h_b$ and $h_e$, both assumed to undergo independent quasi-static fading. The receiver noise variances at Bob and Eve are denoted by $\sn$ and $\se$.

The instantaneous signal-to-noise ratios (SNRs) at Bob and Eve are given by $\gamma_b =\Pt |h_b|^2/\sn$ and $\gamma_e =\Pt |h_e|^2/\se$, respectively, each having an exponential distribution given by
\begin{eqnarray}\label{eq:}
        f_i(\gamma_i) = \frac{1}{\Gi}\exp\left(-\frac{\gamma_i}{\Gi}\right),\,\,\, \gamma_i > 0,\,\,\, i = b, e,
\end{eqnarray}
where $\Gb$ and $\Ge$ are the average SNRs at Bob and Eve, respectively. We assume that Bob and Eve know their individual CSI perfectly and the statistics of both channels are known at Alice. We will consider different assumptions on the availability of Bob's instantaneous CSI at Alice, as discussed later. However, Eve's instantaneous CSI is unknown at Alice, and therefore perfect secrecy cannot always be achieved.

\vspace{-1mm}

\subsection{Existing Secrecy Outage Formulation}

In the existing secrecy outage formulation~\cite{parada_05,bloch_08}, the outage event is defined as $\mathcal{O}(R_s) := \{C_s < R_s\}$, where $R_s>0$ is the target secrecy rate, and\footnote{Here we employ the common notation, $[z]^+ = \max\{0,z\}$.}
\begin{eqnarray}\label{eq:SecCap}
        C_s = [C_b - C_e]^+ = [\log_2(1+\gamma_b) - \log_2(1+\gamma_e)]^+
\end{eqnarray}
is the secrecy capacity, expressed in terms of Bob's channel capacity $C_b$ and Eve's channel capacity $C_e$. The secrecy outage probability is defined as
\begin{eqnarray}\label{eq:Pout}
        p_{\text{out}} = \mathbb{P}(C_s < R_s).
\end{eqnarray}
Here, an outage occurs whenever a message transmission is either unreliable (\ie it cannot be decoded by Bob) or it is not perfectly secure (\ie there is some information leakage to Eve). Whilst this secrecy outage formulation gives a fundamental characterization of the possibility of having a \emph{reliable and secure} transmission, it does not distinguish between reliability and security. Hence, an outage does not necessarily imply a failure in achieving perfect secrecy. Furthermore, if Alice knows that Bob's channel cannot support the secrecy rate, \ie $C_b < R_s$, then she would certainly suspend transmission. This ``suspension event'' falls within the outage event $\mathcal{O}(R_s)$, since $C_b < R_s$ implies $C_s < R_s$; however, it is clearly not a failure in achieving perfect secrecy. From a design perspective, it is important to provide an outage formulation which gives a more explicit measure of the level of security, in order to design transmission schemes which meet target security requirements.

\vspace{-1mm}

\section{Alternative Secrecy Outage Formulation} \label{sec:}

We now present an alternative secrecy outage formulation which directly measures the probability that a \emph{transmitted} message fails to achieve perfect secrecy. Consider the well-known Wyner's encoding scheme~\cite{wyner_75}: The encoder chooses two rates, namely, the rate of the transmitted codewords $R_b$, and the rate of the confidential information $R_s$. The rate difference $R_e \triangleq R_b - R_s$ reflects the cost of securing the message transmission against eavesdropping. For any transmitted message, Bob is able to decode correctly  if $C_b > R_b$, whilst perfect secrecy fails if $C_e > R_e$. Hence, we define the secrecy outage probability as the conditional probability
\begin{eqnarray}\label{eq:Pso}
    \Pins &\triangleq& \mathbb{P}(C_e > R_b - R_s \,|\, \text{message transmission}),
\end{eqnarray}
conditioned upon a message actually being transmitted. Unlike the existing secrecy outage formulation, our new formulation takes into account the system design parameters, such as the rate of the transmitted codewords \emph{as well as} the condition under which message transmissions take place, and therefore provides a more explicit measure of the level of security. When transmission always occurs, which is usually the case if Alice has absolutely no knowledge about the instantaneous CSI of Bob's channel, the secrecy outage probability reduces to the unconditional probability $\mathbb{P}(C_e > R_b - R_s)$, which was considered in~\cite{tang_09}. More generally, when some form of instantaneous CSI of Bob's channel is available, Alice can decide whether or not to transmit with possibly variable rates according to the channel condition. This is important since, as we will see, by carefully designing the condition for transmission, it is possible to dramatically reduce the secrecy outage probability. With this new formulation, the system designer can use the secrecy outage probability to characterize the security level and design transmission schemes that meet target security requirements.

\vspace{-1mm}

\section{Secure Transmission Design} \label{sec:}

In this section, we consider the design problem of maximizing the throughput $\eta=\Ptx R_s$, where $\Ptx$ denotes the probability of transmission. Note that $\Ptx$ can be interpreted as a quality of service (QoS) measure.\footnote{For strictly delay-limited systems, $1-\Ptx$ represents the probability of a message packet being dropped. For systems allowing moderate delays, $p^{-1}_{\text{tx}}-1$ may give an indication of the average delay of transmission.} The primary design requirement is on the security level. In addition, one can impose another constraint on the probability of transmission as a QoS requirement. The design problem can be written as
\begin{eqnarray}\label{eq:OptProbGen}
        \max \,\eta,\,\,\,\, \text{s.t.}\,\, \,  \Pins\leq\epsilon,\; \Ptx\geq\sigma,
\end{eqnarray}
where $\epsilon\in[0,1]$ and $\sigma\in[0,1]$ represent the minimum security and QoS requirements. The controllable design parameters are the rate of the transmitted codewords, $R_b$, the rate of the confidential information, $R_s$, as well as the condition for transmission. In the following, we consider the design of two different transmission schemes, based on Alice having either full or limited knowledge of the instantaneous CSI of Bob's channel. The condition for transmission is given by an on-off scheme; \ie transmission takes place whenever the value of $\gamma_b$ exceeds some SNR threshold $\Gt$. Note that $\Gt\geq 2^{R_s}-1$ always holds, since transmission can only happen when $C_b \geq R_s$.

\vspace{-1mm}

\subsection{Adaptive Encoder with On-Off Transmission} \label{sec:}

First, we consider the scenario where the encoder is able to adaptively choose the rate of the transmitted codewords, $R_b$, according to the instantaneous CSI of Bob's channel. In fact, only the instantaneous SNR, $\gamma_b$, needs to be fed back from Bob to Alice. The encoder sets $R_b$ arbitrarily close to $C_b$, \ie the largest possible rate without incurring any decoding error at Bob. For any given values of $R_s$ and $\Gt$, we can compute the transmission probability as\vspace{-1mm}
\begin{eqnarray}\label{eq:Ptrans2}
        \Ptx = \mathbb{P}(\gamma_b > \Gt) = \exp(-\Gt/\Gb).
\end{eqnarray}
Also, the secrecy outage probability is given by\vspace{-1mm}
\begin{eqnarray}
        \!\!\Pins \!\!\!\!&=&\!\!\!\! \mathbb{P}(C_e > C_b - R_s \, | \, \gamma_b > \Gt)\nonumber\\
         \!\!\!\!&=&\!\!\!\! \mathbb{P}\left(\gamma_b < 2^{R_s}(1+\gamma_e)-1\, | \, \gamma_b > \Gt\right)\nonumber\\
        \!\!\!\!&=&\!\!\!\! \frac{\mathbb{P}\left(\Gt < \gamma_b < 2^{R_s}(1+\gamma_e)-1\right)}
        {\mathbb{P}(\gamma_b>\Gt)}\nonumber\\
        \!\!\!\!&=&\!\!\!\! \exp\left(\!\frac{\Gt}{\Gb}\!\right)\!\!
        \int_{\frac{\Gt+1}{2^{R_s}}-1}^\infty\!\!
        \left(\!\int_{\Gt}^{2^{R_s}(1+\gamma_e)-1}\!\!f_b(\gamma_b) \mathrm{d}\gamma_b       \!\!\right)\!f_e(\gamma_e)\mathrm{d}\gamma_e\nonumber\\
         \!\!\!\!&=&\!\!\!\! \frac{\Ge 2^{R_s}}{\Ge 2^{R_s}+\Gb} \exp\left(-\frac{\Gt+1-2^{R_s}}{\Ge 2^{R_s}}\right)\label{eq:Pint2}.
\end{eqnarray}
We see that the SNR threshold for on-off transmission, $\Gt$, presents a trade-off between the QoS and security. More specifically, a higher security level can be achieved by compromising the QoS, which is done by choosing a larger~$\Gt$.

Now, we consider the design problem of finding the values of $R_s$ and $\Gt$ that maximize the throughput\footnote{Ideally, one can adaptively change $R_s$ according to the instantaneous SNR. However, the optimal solution for this adaptive design is very difficult to obtain. Here, we look for the optimal $R_s$ that is constant over time.}, given by\vspace{-1mm}
\begin{eqnarray}
        \arg \max_{R_s, \Gt} \!\!&&\!\! \Ptx(\Gt) \,R_s, \nonumber\\
        \text{s.t.} \!\!&&\!\! \Pins(\Gt, R_s)\leq\epsilon, \, \Ptx(\Gt)\geq\sigma, \, \Gt\geq 2^{R_s}-1,\, R_s > 0,\nonumber
\end{eqnarray}
where we have explicitly shown the dependence of $\Ptx$ and $\Pins$ on $\Gt$ and $R_s$. To solve this problem, we first establish an expression for the optimal $\Gt$ for any given $R_s$ as follows: Using (\ref{eq:Ptrans2}), the feasible range of $\Gt$ that satisfies $\Ptx(\Gt)\geq\sigma$ is found as $\Gt\in [2^{R_s}\!-\!1,\Gb\ln\sigma^{-1}]$. The throughput maximizing $\Gt$ takes its smallest possible value while satisfying the security constraint. Since $\Pins(\Gt, R_s)$ in (\ref{eq:Pint2}) is a decreasing function of $\Gt$, we obtain the optimal $\Gt$ as\vspace{-1mm}
\begin{eqnarray}
\!\Gt \!=\! \left\{ \begin{array}{ll}\!\!\!
 2^{R_s}\!-\!1, &\mbox{\!\!\!if
 $R_s\!\leq\!\log_2\frac{\Gb\epsilon}{\Ge(1-\epsilon)}$,} \\\!\!\!
  2^{R_s}(1 \!-\! \Ge\ln\epsilon \!+\! \Ge\ln \frac{\Ge 2^{R_s}}{\Ge 2^{R_s}+\Gb})\!-\!1, &\mbox{\!\!\!otherwise,}
       \end{array} \right.\nonumber
\end{eqnarray}
where the first case presents the scenario that setting $\Gt$ to its smallest value still satisfies the security constraint of $\Pins\leq\epsilon$; otherwise, we have the second case by solving $\Pins=\epsilon$.

The condition for having secure communication with some positive rate while satisfying both constraints can be found as
\begin{eqnarray}\vspace{-1mm}
        \epsilon > \frac{\Ge}{\Ge+\Gb}\sigma^{\Gb/\Ge}.\nonumber
\end{eqnarray}
Under this condition, we can formulate the optimization problem as\vspace{-4mm}
\begin{eqnarray}
        \arg \max_{R_s} && R_s \exp(-\Gt/\Gb),\nonumber\\
        \text{s.t.} && R_s > 0,\, \Gt\leq \Gb\ln\sigma^{-1},\nonumber
\end{eqnarray}
where $\Gt$ is a function of $R_s$, whose expression was found above.
Although the dependence of $\Gt$ on $R_s$ makes it difficult to obtain a closed-form solution, this problem can be easily solved numerically.

\vspace{-1mm}

\subsection{Non-Adaptive Encoder with On-Off Transmission} \label{sec:NonAdaOnOff}

Now we consider the transmission design with a non-adaptive encoder in which the rate of the codewords, $R_b$, is constant over time (but needs to be optimally chosen). In contrast to the adaptive encoder design which requires the feedback of the instantaneous SNR, this design only requires 1 bit of feedback to enable the on-off transmission scheme. For any given values of $R_b$, $R_s$, and $\Gt$, the transmission probability is given by (\ref{eq:Ptrans2})
and the secrecy outage probability is given by\vspace{-1mm}
\begin{eqnarray}
        \Pins\!\!\! &=&\!\!\! \mathbb{P}(C_e > R_b - R_s \, | \, \gamma_b > \Gt)
        = \mathbb{P}(C_e > R_b - R_s)\nonumber\\
         \!\!\!&=&\!\!\! \exp\left(-\frac{2^{R_b-R_s}-1}{\Ge}\right)\label{eq:Pint3}.
\end{eqnarray}

\begin{figure}[!t]
\centering\vspace{-3mm}
\includegraphics[width=0.9\columnwidth]{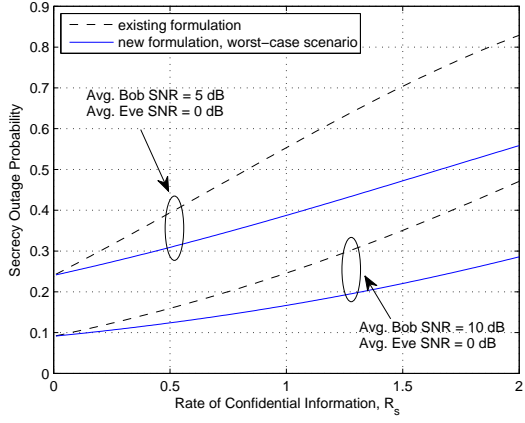}
\vspace{-3mm} \caption{Comparison between existing and new secrecy outage formulations.} \label{fig:outage}\vspace{-2mm}
\end{figure}

Again, we consider the design problem of finding the values of $R_b$, $R_s$ and $\Gt$ that maximize the throughput, given by\vspace{-1mm}
\begin{eqnarray}
        \arg \max_{R_b, R_s, \Gt} \!\!\!\!&&\!\!\!\! \Ptx(\Gt)\,R_s,\nonumber\\ \text{s.t.} \!\!\!\!&&\!\!\!\! \Pins(R_b,R_s)\!\leq\!\epsilon,\, \Ptx(\Gt)\!\geq\!\sigma,\, \Gt\!\geq\! 2^{R_b}\!-\!1,\, R_s \!>\! 0.\nonumber
\end{eqnarray}

We first establish expressions for the optimal $\Gt$ and $R_s$ for any given $R_b$ as follows: Since $\Pins(R_b, R_s)$ in (\ref{eq:Pint3}) is independent of $\Gt$, it is optimal to minimize $\Gt$ in order to maximize $\Ptx(\Gt)$ in (\ref{eq:Ptrans2}). Hence, we obtain the optimal $\Gt$ as\vspace{-1mm}
\begin{eqnarray}\label{eq:}
        \Gt = 2^{R_b}-1.\nonumber
\end{eqnarray}
Note that $R_b \leq\log_2(1+\Gb\ln\sigma^{-1})$ is required or $\Ptx(\Gt)\geq\sigma$ can never be satisfied. For any given $R_b$, the throughput maximizing $R_s$ takes its maximum possible value while satisfying $\Pins(R_b,R_s)\leq\epsilon$. Hence, we obtain the optimal $R_s$ as\vspace{-1mm}
\begin{eqnarray}\label{eq:}
        R_s = R_b - \kappa, \,\,\,\text{where}\,\, \kappa=\log_2(1+\Ge\ln\epsilon^{-1}).\nonumber
\end{eqnarray}
The condition for having secure communication with some positive rate while satisfying both constraints can be found as\vspace{-1mm}
\begin{eqnarray}\label{eq:}
        \epsilon > \sigma^{\Gb/\Ge}.\nonumber
\end{eqnarray}
Under this condition, we can formulate the optimization problem as\vspace{-3mm}
\begin{eqnarray}
        \arg \max_{R_b} && \left(R_b-\kappa\right)
        \exp\left(-\frac{2^{R_b}-1}{\Gb}\right),\nonumber\\
        \text{s.t.} &&\kappa < R_b \leq \log_2(1+\Gb\ln\sigma^{-1}) , \nonumber
\end{eqnarray}
which in this case admits the explicit solution\vspace{-1mm}
\begin{eqnarray}\label{eq:}
        \!R_b \!=\! \min\left\{\!\kappa \!+\! \frac{1}{\ln2}\mathrm{W}_0\Big(\!\Gb\exp(-\kappa\ln2)\!\Big) ,\, \log_2(1\!+\!\Gb\ln\sigma^{-1})\!\right\},\nonumber
\end{eqnarray}
where $\mathrm{W}_0(\cdot)$ is the principal branch of Lambert's ${\rm W}$ function.

\vspace{-1mm}

\section{Numerical Results and Discussion} \label{sec:}

Fig.~\ref{fig:outage} presents the secrecy outage probability, comparing the existing formulation (\ref{eq:Pout}) and the new formulation (\ref{eq:Pint2}). In both cases, the rate of transmitted codewords is adaptively chosen as $R_b = C_b$. For the new formulation, the on-off SNR threshold is set to its minimum value of $\Gt = 2^{R_s}-1$, which gives the maximum outage probability. We see from the figure that the outage probabilities predicted by the two formulations differ significantly. Hence, the existing formulation cannot be directly used to measure the security level.

\begin{figure}[!t]
\centering\vspace{-3mm}
\includegraphics[width=0.9\columnwidth]{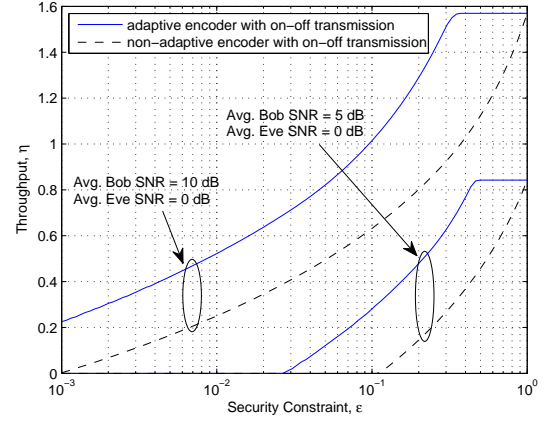}
\vspace{-3mm} \caption{Throughput of secure transmission versus the security constraint. The QoS constraint is fixed to $\sigma=0.5$.} \label{fig:throughput}\vspace{-2mm}
\end{figure}

Fig.~\ref{fig:throughput} compares the achievable throughput of our proposed adaptive and non-adaptive encoder designs, for a range of security constraints. The QoS constraint is fixed to $\sigma=0.5$. We see that it is generally impossible to achieve arbitrarily low secrecy outage probability, which agrees with the lower bounds on the secrecy outage probability derived in Section~IV. This figure also shows that a significant throughput improvement is obtained by adaptively changing the rate of the transmitted codewords based on knowledge of Bob's instantaneous CSI. On the other hand, the design with non-adaptive encoder only requires $1$ bit of feedback, which minimizes the feedback overhead.

\vspace{-1mm}

\bibliographystyle{IEEEtran}


\end{document}